\documentclass[aps,twocolumn]{revtex4}
\usepackage{epsfig}
\usepackage{times}
\usepackage{color}
\begin{document}

\title{Sustainable institutionalized punishment requires elimination of second-order free-riders}

\author{Matja{\v z} Perc}
\email{matjaz.perc@uni-mb.si}
\affiliation{Faculty of Natural Sciences and Mathematics, University of Maribor, Koro{\v s}ka cesta 160, SI-2000 Maribor, Slovenia}

\begin{abstract}
Although empirical and theoretical studies affirm that punishment can elevate collaborative efforts, its emergence and stability remain elusive. By peer-punishment the sanctioning is something an individual elects to do depending on the strategies in its neighborhood. The consequences of unsustainable efforts are therefore local. By pool-punishment, on the other hand, where resources for sanctioning are committed in advance and at large, the notion of sustainability has greater significance. In a population with free-riders, punishers must be strong in numbers to keep the ``punishment pool'' from emptying. Failure to do so renders the concept of institutionalized sanctioning futile. We show that pool-punishment in structured populations is sustainable, but only if second-order free-riders are sanctioned as well, and to a such degree that they cannot prevail. A discontinuous phase transition leads to an outbreak of sustainability when punishers subvert second-order free-riders in the competition against defectors.
\end{abstract}

\maketitle

The provisioning of social benefits or the preservation of environmental resources relies on selfless contributions and collaborative efforts \cite{baumol_52, ostrom_90, gintis_05, sigmund_10}. Those that exploit such public goods are therefore faced with individuals and institutions that are prepared to sanction antisocial behavior \cite{fehr_aer00, fehr_n02, gardner_a_an04, henrich_s06b, rockenbach_n06, henrich_s06, gurerk_s06, brandt_pnas06, gachter_s08, rockenbach_n09} with the aim of averting Hardin's tragedy of the commons \cite{hardin_g_s68}. The Achilles' heel of punishment, however, is the fact that it is costly, weighing heavily on the shoulder of those that already fill the common pool \cite{fehr_n03, boyd_pnas03, fowler_pnas05, milinski_n08, eldakar_pnas08}. In the presence of punishers, the traditional cooperators, i.e., those that contribute to the public good but do not punish, are therefore downgraded to free-riders as well. This so-called second-order free-riding is in many ways more prohibitive for the emergence and stability of punishment then the traditional defectors \cite{panchanathan_n04, fehr_n04, fowler_n05b}. Without additional incentives and mechanisms that help sustain punishment, the second-order free-riders prevail over punishers, thereby eliminating the threat of sanctioning. In well-mixed populations, volunteering \cite{hauert_s02, semmann_n03, mathew_prsb09} can cause this unfortunate scenario to unravel \cite{hauert_s07}, as can coordinated efforts between the punishers \cite{boyd_s10}. However, spatial structure, in contrast to well-mixed interactions, may alone be sufficient to solve the second-order free-rider problem \cite{helbing_ploscb10, helbing_njp10}.

In spite of the predominantly positive acclaim, studies critically probing the effectiveness of punishment in promoting collaborative efforts among unrelated and selfish individuals, for example in conjunction with anti-social punishment \cite{herrmann_s08, rand_jtb10, rand_nc11}, indirect reciprocity \cite{ohtsuki_n09} or social differences \cite{wu_jj_pnas09, gintis_n08}, are an important reminder of open questions still imbuing the subject. As a viable alternative to punishment, rewarding has received substantial attention as well \cite{sigmund_pnas01, andreoni_aer03, hilbe_prsb10, szolnoki_epl10}. Although the majority of previous studies addressing the ``stick versus carrot'' dilemma concluded that punishment is more effective than reward in sustaining public cooperation \cite{sigmund_tee07}, evidence suggesting that rewards may be as effective as punishment and lead to higher total earnings without potential damage to reputation \cite{milinski_n02, semmann_bes04} or fear from retaliation are mounting \cite{dreber_n08, rand_s09}.

Here we also depart from the traditional model of punishment by considering it not as an individually inspired act, i.e., peer-punishment, but rather as something that is inherent to the population as a whole, i.e., pool-punishment. Although both variants entail paying a cost for the free-riders to incur a cost, by peer-punishment this is done after the public goods game and is primarily an individually inspired effort, while by pool-punishment contributions to the ``punishment pool'' are summoned in advance. As pointed out in a recent study by Sigmund et al. \cite{sigmund_n10}, the first experiment on public goods with punishment \cite{yamagishi_jpsp86} actually considered pool- rather than peer-punishment. An important advantage of collecting resources for punishment ahead of the collaborative effort, as in paying towards a sanctioning institution, is the fact that second-order free-riders are easily spotted and are thus submissive to being punished. Note that if everyone contributes to the public good then second-order free-riders are not distinguishable from peer-punishers. Pool-punishment alleviates this important deficiency, and as a reported in \cite{sigmund_n10}, can prevail over peer-punishment if second-order free-riders are punished as well.

Our model is based on the spatial public goods game (see e.g. \cite{szolnoki_pre09c, gomez-gardenes_epl11, gomez-gardenes_c11, szolnoki_pre11c, szolnoki_pre11b}) and entails punishers (P), cooperators (C) and defectors (D) as the three strategies competing for dominance on the square lattice. Notably, since we consider structured rather than well-mixed populations, the option of volunteering \cite{hauert_s07} for stabilizing either cooperation or punishment is not required \cite{helbing_ploscb10}. Punishers and cooperators both contribute equally to the public good. The resulting amount is multiplied by the synergy factor $r>1$ and then divided equally among the group members irrespective of their strategies. But while punishers contribute an amount $\beta$ also to the punishment pool, the cooperators refrain from doing the same. Due to their second-order free riding the cooperators are fined an amount $\delta \gamma$, where $\delta \leq 1$ is a multiplication factor taking into account the fact that their offence is lesser than the one committed by defectors. The latter contribute neither to the public good nor to the punishment pool and are therefore charged for the full amount $\gamma$.

Since institutions for governing the commons supposedly act as the ``invisible hand'' looming over the whole population, the free-riders are finned irrespective of their neighborhoods, in particular, irrespective of whether they contain a punisher or not. Likewise, contributions to the punishment pool are summoned irrespective of the number of free-riders in the population. This is an important distinction from previously studied spatial public goods games incorporating peer-punishment \cite{brandt_prsb03, nakamaru_eer05, sekiguchi_jtb09, helbing_pre10c}, where punishing costs and fines were deducted from payoffs individually on the basis of strategies that were present in a particular group at a given time. The non-local character of pool-punishment allows for the introduction of an account balance $\Sigma$ for the punishment pool. If the contributions of all the punishers cover the costs that are need to punish all the free-riders in the population, i.e., if $\Sigma \geq 0$, the pool-punishment is said to be fully sustainable. Conversely, if $\Sigma < 0$ the pool-punishment is unsustainable. Sustainability is key for every institution to remain in existence, and sanctioning institutions should be no exception to this assertion. In exceptional cases even a small amount of punishers, although technically unable to sustain $\Sigma \geq 0$, can still ensure enough resources to punish free-riders, for example by means of lobbying or loans and similar financial mechanisms. Such situations can be dubbed accordingly as being conditionally sustainable, and we will make note of them when presenting the results.

Beforehand, our main discoveries for a society facing public goods games with pool-punishment may be summarized as follows. First, we show that the spatial structure can resolve the second-order free-rider problem in case of institutionalized punishment. Without any additional assumptions or strategic complexity, pool-punishers can fully eliminate cooperators. This happens by means of a discontinuous phase transition leading to an outbreak of sustainability, either full or conditional, depending further on the punishment fine and the synergetic effects of collaborative efforts. Second, sustainable pool-punishment is possible exclusively if second-order free-riders go extinct. Only beyond the discontinuous phase transition, by means of which cooperators are eliminated, can the punishers keep the punishment pool from emptying and maintain a positive balance. Importantly though, the elimination of second-order free-riders is only the necessary but not the sufficient criterion for full sustainability of pool-punishment. For small punishment fines and modest synergetic effects of collaborative efforts the defectors can still overburden the sanctioning institution, leading to a conditionally sustainable state only. Remarkable nevertheless is the fact that by appropriate conditions up to half of the population may be defectors, and still the pool-punishment remains fully sustainable. This is in sharp contrast with the fact that even a minute fraction of cooperators precludes sustainable institutionalized punishment, and strengthens the perception that not the defectors but rather the second-order free-riders are the ones that compromise the success of punishment the most.

\begin{figure*}
\begin{center}
\includegraphics[width=10.335cm]{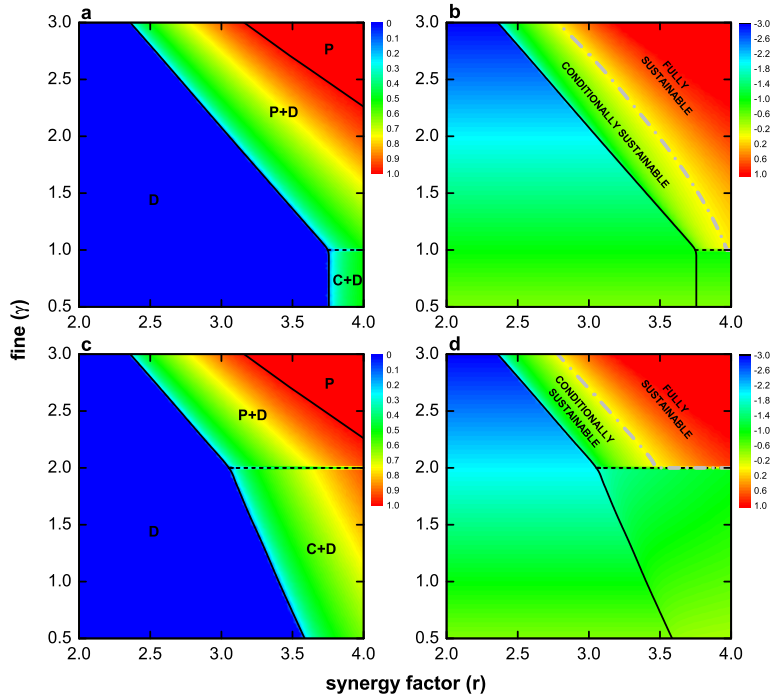}
\end{center}
\caption{\textbf{Phase diagrams (a,c) of the spatial public goods game with punishers (P), cooperators (C) and defectors (D), and the resulting pool balance $\Sigma$ (b,d) on the $r-\gamma$ parameter plane.} In (a) and (b) $\delta=1.0$ while in (c) and (d) $\delta=0.5$. Black solid (dashed) lines depict continuous (discontinuous) phase transitions. Color mapping in the phase diagrams (a,c) encodes the density of punishers in the P+D region and the density of cooperators in the C+D region. Pure P and D phases are depicted red and blue, respectively. In (b) and (d) the color map encodes the punishment pool balance $\Sigma$ pertaining to the phase diagrams on the left. The phase separation lines are depicted for reference as well. Gray dash-dotted lines delineate the region of full sustainability, i.e., $\Sigma \geq 0$, while the region where $\rho_{\rm P}>0$ and $\Sigma < 0$ is denoted as conditionally sustainable. Irrespective of $\delta$, sufficiently large $r$ and $\gamma$ can stabilize pool-punishment by means of a discontinuous phase transition at which punishers replace cooperators in the coexistence with defectors (${\rm C}+{\rm D} \to {\rm P}+{\rm D}$) or via a continuous phase transition where ${\rm D \to P+D}$. If second-order free-riders are punished equally strong as defectors (a,b) the critical fine at which the discontinuous phase transition occurs is smaller than if cooperators are finned only half as strong as defectors (c,d). Accordingly, in (a) and (b) both sustainability regions extend further towards smaller $\gamma$, albeit shifting towards ever larger $r$ as well. The emergence of the P+D phase shifts the sustenance of collaborative efforts toward smaller $r$ as $\gamma$ increases. This can be observed best for $\delta=1.0$, where the punishment of cooperators and defectors is equally strong, and hence in the absence of punishers their coexistence is independent of $\gamma$. However, as soon as punishers subvert cooperators at $\gamma=1.0$ by means of a discontinuous phase transition, the cooperative behavior starts existing also for $r \leq 3.74$. Qualitatively identical features can be observed for $\delta=0.5$ (c,d). Results in all panels were obtained for $\beta=1.0$ and $K=0.5$.}
\label{fig1}
\end{figure*}

\section*{Results}
In the absence of punishment cooperators survive only if $r>3.74$, and crowd out defectors completely for $r>5.49$ \cite{szolnoki_pre09c}. These can be used as benchmark values for evaluating the impact of pool-punishment on the evolution of cooperation in structured populations.

Focusing first on the $r - \gamma$ parameter plane, we present in Fig.~\ref{fig1} full phase diagrams and the corresponding dependence of the punishment pool balance $\Sigma$ for two different values of $\delta$. Panels (a) and (b) feature results for $\delta=1.0$, implying that second-order free-riders are punished equally strong as defectors. This is a common assumption, although the offence committed by cooperators, who contribute to the public good but abstain from punishing, is actually smaller than the one committed by defectors, who free-ride on both occasions. It can be observed that below a critical fine $\gamma=1.0$ punishers cannot survive, and accordingly, the spatial grid is dominated by defectors or a mixed C+D phase, depending on the value of $r$. Since cooperators and defectors are punished equally, the absence of punishers transforms the public goods game into its traditional two-strategy variant where the value of $\gamma$ merely rescales the payoffs but does not influence the outcome. In this sense, it could be assumed that punishment is not executed at all if punishers die out without this affecting the presented results. Accordingly, $r=3.74$ \cite{szolnoki_pre09c} is recovered as the critical synergy factor above which cooperators can survive. For $\gamma>1.0$, however, cooperators are subverted by punishers via a discontinuous phase transition. With their emergence, collaborative efforts can be sustained also for $r<3.74$, whereby the larger the fine the smaller the synergy factor needed to achieve this. Pertaining variations in sustainability, evaluated by means of the punishment pool balance $\Sigma$, are depicted in panel (b). The elimination of second-order free-riders is clearly a necessary condition that needs to be fulfilled for pool-punishment to be sustainable. In addition, however, the fraction of defectors in the P+D phase needs to be sufficiently small. The region of full sustainability, where $\Sigma>0$, is delineated with a dash-dotted gray line, while conditional sustainability characterizes the remainder of the P+D phase. There the high fraction of defectors in the P+D phase precludes positive values of $\Sigma$.

\begin{figure*}
\begin{center}
\includegraphics[width=16.00cm]{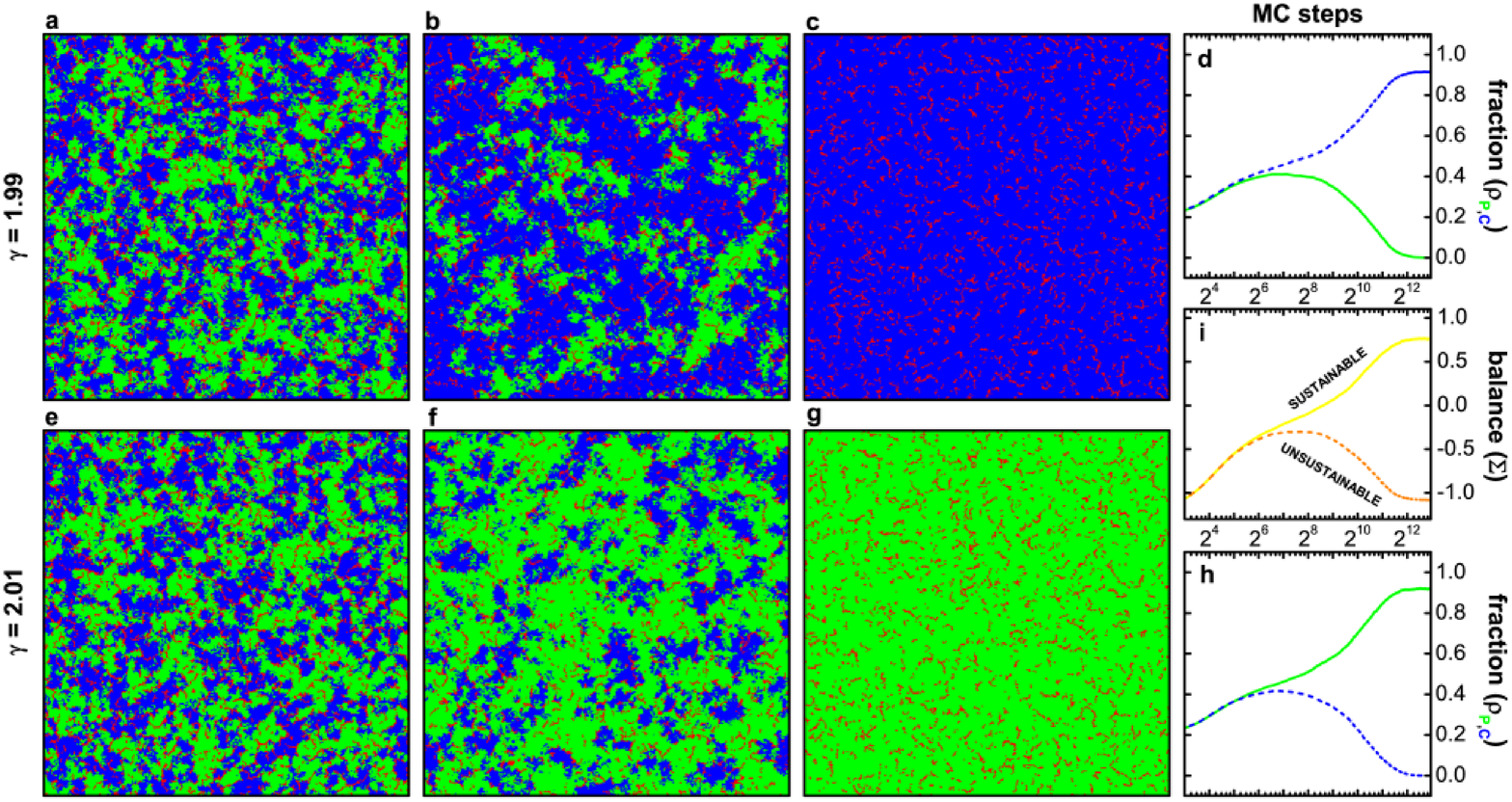}
\end{center}
\caption{\textbf{Characteristic snapshots of the square lattice and time courses of strategy densities close to the discontinuous phase transition at $\gamma=2.0$.} Panels (a), (b) and (c), obtained for $\gamma=1.99$, depict snapshots leading to the coexistence of cooperators (blue) and defectors (red). Time courses in (d) show the pertaining elimination of punishers (green) at the expense of second-order free-riders (dashed blue). Just on the other side of the discontinuous phase transition, at $\gamma=2.01$, panels (e), (f) and (g) depict snapshots leading to the coexistence of punishers (green) and defectors (red). Accordingly, the two time courses in (h) show the pertaining elimination of second-order free-riders (dashed blue) at the expense of punishers (green). Panel (i) depicts the time evolution of the punishment pool balance $\Sigma$ for $\gamma=1.99$ (dashed orange) and $\gamma=2.01$ (yellow) towards unsustainability ($\Sigma < 0$) and sustainability ($\Sigma > 0$), respectively. Snapshots were taken at $100$ (a,e), $1000$ (b,f) and $10000$ (c,h) full Monte Carlo (MC) steps. Note that at $\pm 0.01$ distance in the value of $\gamma$ from the discontinues phase transitions the system can actually be considered as being far away from it. Setting $\gamma$ an order of magnitude closer to the transition point would prolong the equilibration time dramatically. The horizontal axis in (d,i,h) is logarithmic. Results in all panels were obtained for $r=4.0$, $\delta=0.5$, $\beta=1.0$ and $K=0.5$.}
\label{fig2}
\end{figure*}

Complying with the proposition that cooperators should be punished more leniently than defectors, we set $\delta$ to $0.5$, which implies that the fine for second-order free-riding is half the one for defecting. Results presented in panels (c) and (d) of Fig.~\ref{fig1} qualitatively agree with those presented for $\delta=1.0$ in panels (a) and (b). A distinctive feature is that the critical fine at which the discontinuous phase transition ${\rm C}+{\rm D} \to {\rm P}+{\rm D}$ occurs is now two times larger, i.e., $\gamma=2.0$. Borders of conditional and full sustainability move towards larger $\gamma$ accordingly. It is important to note that in the absence of cooperators $\delta$ has no effect on the evolution of the remaining two strategies. For $\gamma>2.0$ the results in panels (c) and (d) are therefore identical to those presented in panels (a) and (b). Conversely, for $\gamma<2.0$, where punishers cannot survive, the competition between cooperators and defectors is no longer unaffected by $\gamma$ as was this the case for $\delta=1.0$. Since here defectors are punished with the full fine while cooperators with only half of it, larger values of $\gamma$ decrease the critical synergy factor $r$ that is needed for cooperative behavior to remain in existence.

Since the discontinuous phase transition ${\rm C}+{\rm D} \to {\rm P}+{\rm D}$ is crucial for the sustainability of pool-punishment, we proceed by examining the evolutionary process at both sides of it in detail. Panels (a), (b) and (c) of Fig.~\ref{fig2} depict characteristic snapshots of the spatial grid that eventually lead to a stable C+D phase, while panels (e), (f) and (g) depict snapshots leading to a stable P+D phase. Although the final outcome in this two cases is remarkably different, the difference in the fine $\gamma$ that is needed for this to happen is minute, which is a characteristic feature of a discontinuous phase transition. Results in panels (d) and (h) illustrate this phenomenon in terms of the densities of punishers $\rho_{\rm P}$ (solid green) and cooperators (dashed blue) $\rho_{\rm C}$ over time. What promises to be an identical evolutionary process at $100$ full Monte Carlo (MC) iteration steps [compare (a) and (e)], \textit{slowly} diverges [see (d) and (h)] towards two very different but stable results. At $1000$ full MC steps the snapshots already hint decisively in favor of either the cooperators (b) or the punishers (f), depending on the value of $\gamma$. At $10000$ full MC steps the stationary state in both cases is reached, where the full magnitude of the difference is revealed. For $\gamma=1.99$ (c) the cooperators completely subvert punishers to form a stable coexistence phase with defectors (red), while for $\gamma=2.01$ (g) the punishers prevail and eliminate the second-order free-riders completely. Panel (i) illustrates the workings of the discontinuous phase transition in terms of the punishment pool balance $\Sigma$, which, after a substantial period of equivalence, turns unsustainable for $\gamma=1.99$ (dashed orange) and sustainable for $\gamma=2.01$ (solid yellow).

\begin{figure*}
\begin{center}
\includegraphics[width=10.092cm]{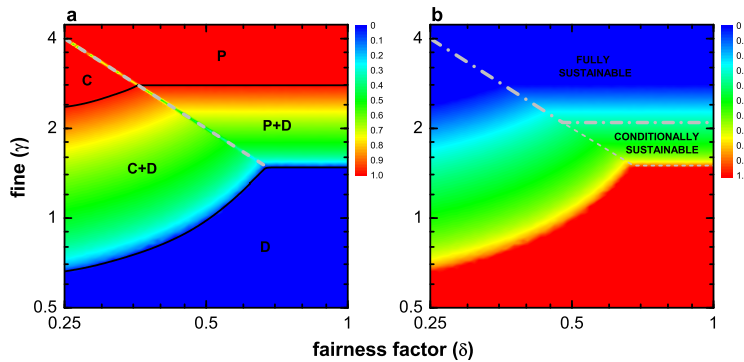}
\end{center}
\caption{\textbf{Phase diagram (a) of the spatial public goods game with punishers (P), cooperators (C) and defectors (D), and the corresponding density of defectors (b) on the $\delta-\gamma$ parameter plane.} In (a) black solid lines depict continuous phase transitions. Dashed gray line depicts the analytically predicted $\gamma=\beta/\delta$ discontinuous phase transition line, which agrees perfectly with the numerical results. Color mapping encodes the density of punishers in the P+D region and the density of cooperators in the C+D region. Pure P and D phases are depicted red and blue, respectively. In (b) the color map encodes the stationary density of defectors pertaining to the phase diagram on the left. Gray dash-dotted line delineates the region of full sustainability, i.e., $\Sigma \geq 0$, while the region where $\rho_{\rm P}>0$ and $\Sigma < 0$ is denoted as conditionally sustainable and delineated with dashed gray. Note that defectors need not die out completely for pool-punishment to be fully sustainable. Remarkably, in the absence of second-order free-riders as much as half of the lattice may still be occupied by defectors when $\Sigma \geq 0$. Results were obtained for $r=3.4$, $\beta=1.0$ and $K=0.5$.}
\label{fig3}
\end{figure*}

Different outcomes of the proposed spatial public goods game with pool-punishment can be understood better still if considering $\delta$ and $\gamma$ as the two variable parameters by a given value of $r$. Figure~\ref{fig3} features the full $\delta - \gamma$ phase diagram (a) and the corresponding color encoded stationary fraction of defectors $\rho_{\rm D}$ (b) for $r=3.4$. The phase diagram has a markedly webbed structure, indicating the possibility of stable pure P, C and D phases, as well as stable mixed P+D and C+D phases. All but the ${\rm C}+{\rm D} \to {\rm P}+{\rm D}$ phase transition are continuous, as indicated by the black solid lines. The discontinuous phase transition is depicted dashed gray, whereby the line corresponds exactly to $\delta=\beta/\gamma$, which can be obtained by equating $P_{\rm P}$ and $P_{\rm C}$ (see Methods). Thus, it implies equivalence of punishers and cooperators. Above the line punishers should outperform cooperators, while below the line cooperators should prevail. Due to the locally independent introduction of pool-punishment, this well-mixed approximation is accurately reproduced by the numerical simulations of the spatially structured model. Panel (b) features the pertaining regions of conditional (dashed gray) and full (dash-dotted gray) sustainability. For sufficiently large (small) values of $\gamma$ ($\delta$) the $\delta=\beta/\gamma$ dependence agrees perfectly with the emergence of full sustainability, and thus confirms that the elimination of second-order free-riders is a necessary condition for institutionalized punishment to be sustainable. Before reaching the ${\rm P}+{\rm D} \to {\rm D}$ phase transition line from above, the pool-punishment becomes conditionally sustainable only ($\Sigma<0$ and $\rho_{\rm P} > 0$), whereby the discontinuous phase transition on the left still remains an accurate delineator of this region. The color encoded values of $\rho_{\rm D}$ in (b) illustrate that under appropriate conditions up to half of the lattice my be occupied by defectors and still $\Sigma$ remains positive. This is in stark contrast with the fact that even a minute fraction of cooperators precludes the possibility of sustainable institutionalized punishment, and leads to the conclusion that not the defectors but rather the second-order free-riders are the ones most prohibitive for its success.

\section*{Discussion}
Over-fishing, environmental pollution, depletion of natural resources, or the misuse of social security systems, are prime examples of the exploitation of public goods. We as humans, although being notoriously famous for cooperative behavior, are also likewise famous for exceeding resources that are readily available to us, despite the fact that later generations may pay for our misbehavior greatly. Identifying what works most efficiently in leading us away from antisocial behavior is therefore of the outmost importance. Punishment, either peer or pool based, is weaved in our history records as something that can foster collaborative efforts and keeps our egos in check. However, in the light of pro-social behavior, punishment can be regarded as just another public good that needs our selfless side to shine through. How and why punishment emerges and can be stabilized appears therefore to be a translation of above-mentioned problems into a single one, which if effectively solved, will solve also all the other ones. Yet it is a fruitful and gratifying approach allowing us to capture the essence of the problem and investigate, primarily by means of minimalist but relevant models, the hows and whys of the evolution of cooperation and social norms \cite{axelrod_84, skyrms_96, henrich_04}.

Here we have demonstrated that such a model can explain the emergence and stability of institutionalized punishment. In particular, we have shown that the elimination of second-order free-riders through spatially structured interactions paves the way for sustainable pool-punishment if accompanied by sufficiently large fines and synergetic effects of cooperation. Second-order free-riders are thereby eliminated by means of a discontinuous phase transition that shifts the evolution rather explosively in favor of the punishers. Although this discontinuity is due to the simplified assumption of unconditional punishment, and partially contradicts with real-life experience in that it prohibits a stable coexistence of second-order free-riders and punishers, it nevertheless outlines a succinct and viable solution of the second-order free-rider problem that is in line with recent advances \cite{sigmund_n10}. Altogether, the presented results strengthen the established importance of the spatial structure in promoting collaborative efforts \cite{nowak_n92b, nowak_pnas94, hauert_n04} as well as in stabilizing punishment \cite{brandt_prsb03, helbing_ploscb10}, and suggest that elaborating further on the particularities of pool-punishment in structured populations, especially with methods of statistical physics \cite{szabo_pr07, castellano_rmp09}, may improve our understanding of the origin of institutions.

\section*{Methods}
The public goods game is staged on a square lattice with periodic boundary conditions. Players play the game with their $k=4$ nearest neighbors. Accordingly, each individual belongs to five different groups containing five players each. Initially each player on site $x$ is designated either as a punisher ($s_x = {\rm P}$), cooperator ($s_x = {\rm C}$) or defector ($s_x = {\rm D}$) with equal probability. Using standard parametrization, the two cooperating strategies P and C contribute $1$ to the public good while defectors contribute nothing. The sum of all contributions in each group is multiplied by the factor $r>1$, reflecting the synergetic effects of cooperation, and the resulting amount is equally divided among the $k+1$ members irrespective of their strategies.

Pool-punishment requires allocating resources by means of which free-riders can subsequently be punished. Each punisher therefore contributes an amount $\beta$ to the punishment pool that is subtracted from its payoff. Since the resources for pool-punishment are actually committed before the collaborative effort, both free-rider types are exposed. Cooperators, whose second-order free-riding can stay undetected by peer-punishment, are spotted just as readily as the defectors. The amount withheld from the common pool by defectors is, however, larger than the one withheld by cooperators. To take this into account and enable ``fair punishing'', each defector is punished with a full fine $\gamma$, while the fine for second-order offenders is reduced by a multiplication factor $\delta \leq 1$. Denoting the number of punishers (P), cooperators (C) and defectors (D) in a given group $g$ by $N_{\rm P}$, $N_{\rm C}$ and $N_{\rm D}$, respectively, the payoffs
\begin{eqnarray}
P_{\rm P}^g&=&[r(N_{\rm P}+N_{\rm C})-\beta]/(k+1) - 1, \nonumber \\
P_{\rm C}^g&=&[r(N_{\rm P}+N_{\rm C})-\delta \gamma]/(k+1) - 1 \,\,\,\,\,{\rm and} \nonumber \\
P_{\rm D}^g&=&[r(N_{\rm P}+N_{\rm C}) - \gamma]/(k+1) \nonumber
\end{eqnarray}
are obtained by each player $x$ depending on its strategy $s_x$. Since pool-punishment corresponds to an institutionalized system that operates population-wide irrespective of local considerations, the costs and fines are subtracted from the appropriate players irrespective of their neighborhoods. This \textit{unconditional} execution of punishment takes into account the ``invisible hand'' of justice looming over the free-riders. Although being a simplification (for an alternative treatment see \cite{szolnoki_pre11}), it accounts for the fact that the same effect is missing by peer-punishment, where the threatening omnipresence of a sanctioning institution is absent and the execution of punishment is therefore neighborhood-dependent. However, since punishers contributing to the punishment pool may not be strong enough in numbers to actually gather enough resources to punish all the free-riders in a population, the necessity to assess the sustainability of pool-punishment emerges. The account balance $\Sigma = \beta \rho_{\rm P}  - \gamma (\delta \rho_{\rm C}  + \rho_{\rm D})$ of the punishment pool is thus defined, where $\rho_{s_x}$ are the stationary fractions of strategies on the $L \times L$ square lattice. If $\Sigma \geq 0$ the pool has a zero or positive balance, and thus the pool-punishment is said to be sustainable. Conversely, if $\Sigma < 0$ the resources needed to execute punishment exceed the contributions to the pool, and accordingly, the pool-punishment is unsustainable. It is also possible to argue that even a small amount of punishers can still ensure enough resources to punish free-riders. Situations where $\rho_{\rm P} > 0$ and $\Sigma < 0$ can be referred to appropriately as being conditionally sustainable.

The stationary fractions of punishers $\rho_{\rm P}$, cooperators $\rho_{\rm C}$ and defectors $\rho_{\rm D}$ on the square lattice are determined by means of a random sequential update comprising the following elementary steps. First, a randomly selected player $x$ plays the public goods game with its $k$ interaction partners as a member of all the $g=1, \ldots, 5$ groups it belongs to. The overall payoff it thereby obtains is thus $P_{s_x} = \sum_g P_{s_x}^g$. Next, one of the four nearest neighbors of player $x$ is chosen randomly, and its location is denoted by $y$. Player $y$ also acquires its payoff $P_{s_y}$ identically as previously player $x$. Finally, player $y$ imitates the strategy of player $x$ with the probability $q=1/\{1+\exp[(P_{s_y}-P_{s_x})/K]\}$, where $K$ determines the level of uncertainty by strategy adoptions or its inverse $K^{-1}$ the so-called intensity of selection \cite{szabo_pre98}. In the $K \to 0$ limit player $y$ unconditionally imitates player $x$ if $P_{s_y} > P_{s_x}$. Conversely, in the $K \to \infty$ limit all information about the payoffs is lost and player $y$ changes its strategy by means of a coin toss. Without the loss of generality we set $K=0.5$ \cite{szolnoki_pre09c}, implying that better performing players are readily imitated, but it is not impossible to adopt the strategy of a player performing worse. Such errors in judgment can be attributed to mistakes and external influences that affect the evaluation of the opponent. Each full Monte Carlo (MC) step involves all players having a chance to adopt a strategy from one of their neighbors once on average. Depending on the proximity to phase transition points, the linear system size varied from $L=200$ to $1600$ and the equilibration required up to $10^6$ full MC steps for the finite size effects to be avoided.

\noindent \\ \textbf{Acknowledgments} \\
\noindent This research was supported by the Slovenian Research Agency (Grant J1-4055).

\noindent \\ \textbf{Competing financial interests} \\
The author declares no competing financial interests.

\end{document}